\ifx\pdfoutput\undefined
	\documentclass[dvips,apjl]{emulateapj}
	\DeclareGraphicsExtensions{.eps}
\else
	\documentclass[pdftex,apjl]{emulateapj}
  \DeclareGraphicsExtensions{.pdf}
\fi

\usepackage{apjfonts}

\begin{document}

\title{GRB 050315: A step toward the uniqueness of the overall GRB structure}

\author{Remo Ruffini\altaffilmark{1,2}, Maria Grazia Bernardini\altaffilmark{1,2}, Carlo Luciano Bianco\altaffilmark{1,2}, Pascal Chardonnet\altaffilmark{1,3}, Federico Fraschetti\altaffilmark{1,4}, Roberto Guida\altaffilmark{1,2}, She-Sheng Xue\altaffilmark{1,2}}

\altaffiltext{1}{ICRANet and ICRA, Piazzale della Repubblica 10, I--65100 Pescara, Italy. E-mails: ruffini@icra.it, xue@icra.it}
\altaffiltext{2}{Dipartimento di Fisica, Universit\`a di Roma ``La Sapienza'', Piazzale Aldo Moro 5, I-00185 Roma, Italy. E-mails: maria.bernardini@icra.it, bianco@icra.it}
\altaffiltext{3}{Universit\'e de Savoie, LAPTH - LAPP, BP 110, F-74941 Annecy-le-Vieux Cedex, France. E-mail: chardon@lapp.in2p3.fr}
\altaffiltext{4}{Osservatorio Astronomico di Brera, via Bianchi 46, I-23807 Merate (LC), Italy. E-mail: fraschetti@icra.it}

\begin{abstract}
Using the \emph{Swift} data of GRB 050315, we progress on the uniqueness of our theoretically predicted Gamma-Ray Burst (GRB) structure as composed by a proper-GRB (P-GRB), emitted at the transparency of an electron-positron plasma with suitable baryon loading, and an afterglow comprising the so called ``prompt emission'' as due to external shocks. Thanks to the \emph{Swift} observations, the P-GRB is identified and for the first time we can theoretically fit detailed light curves for selected energy bands on a continuous time scale ranging over $10^6$ seconds. The theoretically predicted instantaneous spectral distribution over the entire afterglow is presented, confirming a clear hard-to-soft behavior encompassing, continuously, the ``prompt emission'' all the way to the latest phases of the afterglow.
\end{abstract}


\section{Introduction}

GRB 050315 \citep{va05} has been triggered and located by the BAT instrument \citep{b04,ba05} on board of the {\em Swift} satellite \citep{ga04} at 2005-March-15 20:59:42 UT \citep{pa05}. The narrow field instrument XRT \citep{bua04,bua05} began observations $\sim 80$ s after the BAT trigger, one of the earliest XRT observations yet made, and continued to detect the source for $\sim 10$ days \citep{va05}. The spectroscopic redshift has been found to be $z = 1.949$ \citep{kb05}.

We present here the results of the fit of the \emph{Swift} data of this source in $5$ energy bands in the framework of our theoretical model \citep[see][and references therein]{rlet1,rlet2,rubr,rubr2,EQTS_ApJL,EQTS_ApJL2,PowerLaws}, pointing out a new step toward the uniqueness of the explanation of the overall GRB structure. In section \ref{model} we recall the essential features of our theoretical model; in section \ref{fit} we fit the GRB 050315 observations by both the BAT and XRT instruments; in section \ref{spectra} we present the instantaneous spectra for selected values of the detector arrival time ranging from $60$ s (i.e. during the so called ``prompt emission'') all the way to $3.0\times 10^4$ s (i.e. the latest afterglow phases); in section \ref{concl} we present the conclusions.

\section{Our theoretical model}\label{model}

A major difference between our theoretical model and the ones in the current literature \citep[see e.g.][and references therein]{p04} is that what is usually called ``prompt emission'' in our case coincides with the peak of the afterglow emission and is not due to the prolonged activity of an ``inner engine'' which, clearly, would introduce an additional and independent physical process to explain the GRB phenomenon \citep{rlet2}. A basic feature of our model consists, in fact, in a sharp distinction between two different components in the GRB structure: {\bf 1)} the Proper-GRB (P-GRB), emitted at the moment of transparency of the self-accelerating $e^\pm$-baryons plasma \citep[see e.g.][]{g86,p86,sp90,psn93,mlr93,gw98,rswx99,rswx00,rlet1,rlet2,Monaco_RateEq}; {\bf 2)} an afterglow described by external shocks and composed of three different regimes \citep[see][and references therein]{rswx99,rswx00,rlet2,rubr}. The first afterglow regime corresponds to a bolometric luminosity monotonically increasing with the photon detector arrival time, corresponding to a substantially constant Lorentz gamma factor of the accelerated baryons. The second regime consists of the bolometric luminosity peak, corresponding to the ``knee'' in the decreasing phase of the baryonic Lorentz gamma factor. The third regime corresponds to a bolometric luminosity decreasing with arrival time, corresponding to the late deceleration of the Lorentz gamma factor. In some sources the P-GRB is under the observability threshold. In \citet{rlet2} we have chosen as a prototype the source GRB 991216 which clearly shows the existence of the P-GRB and the three regimes of the afterglow. Unfortunately, data from BATSE existed only up to $ 36 $ s, and data from R-XTE and Chandra only after $ 3500 $ s, leaving our theoretical predictions in the whole range between $ 36 $ s and $ 3500 $ s without the support of the comparison with observational data. Nevertheless, both the relative intensity of the P-GRB to the peak of the afterglow in such source, as well as their corresponding temporal lag, were theoretically predicted within a few percent (see Fig. 11 in \citet{rubr}).

\begin{figure*}
\includegraphics[width=0.5\hsize,clip]{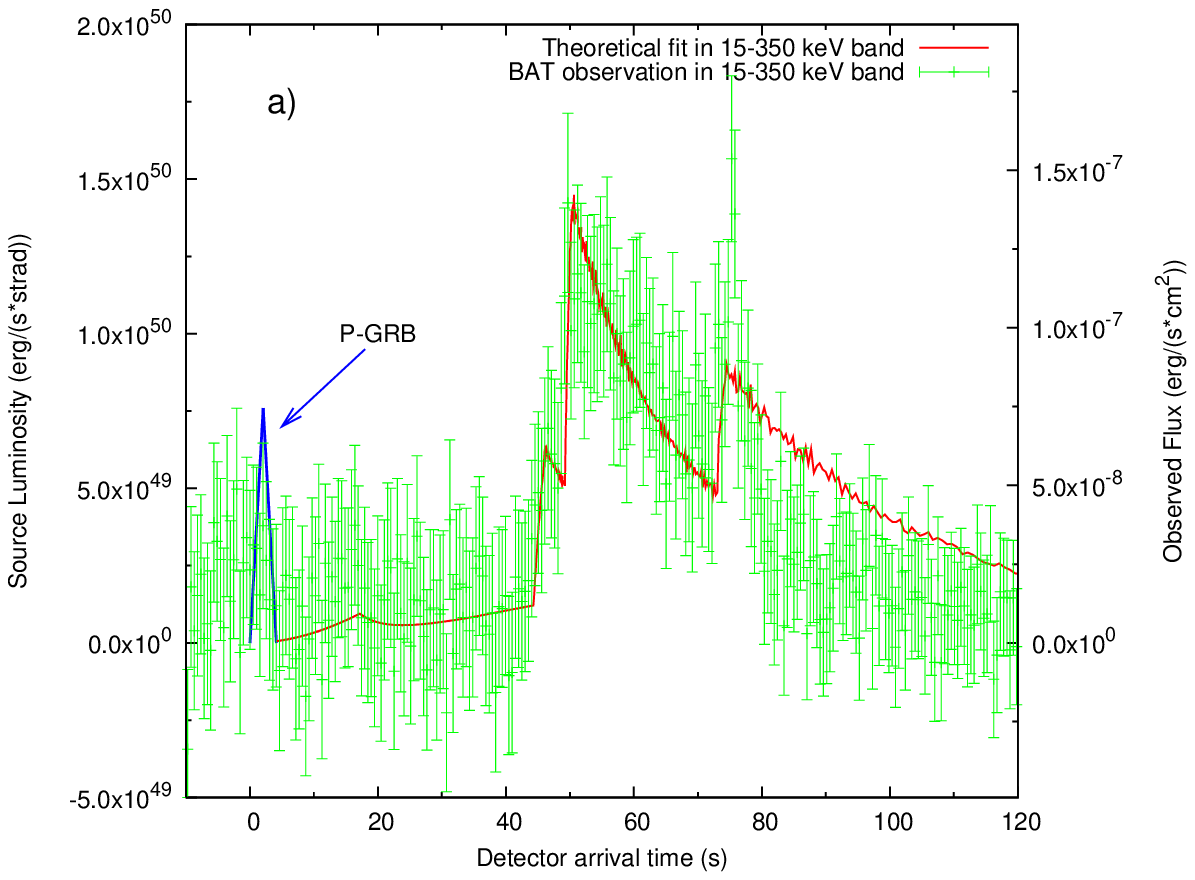}
\includegraphics[width=0.5\hsize,clip]{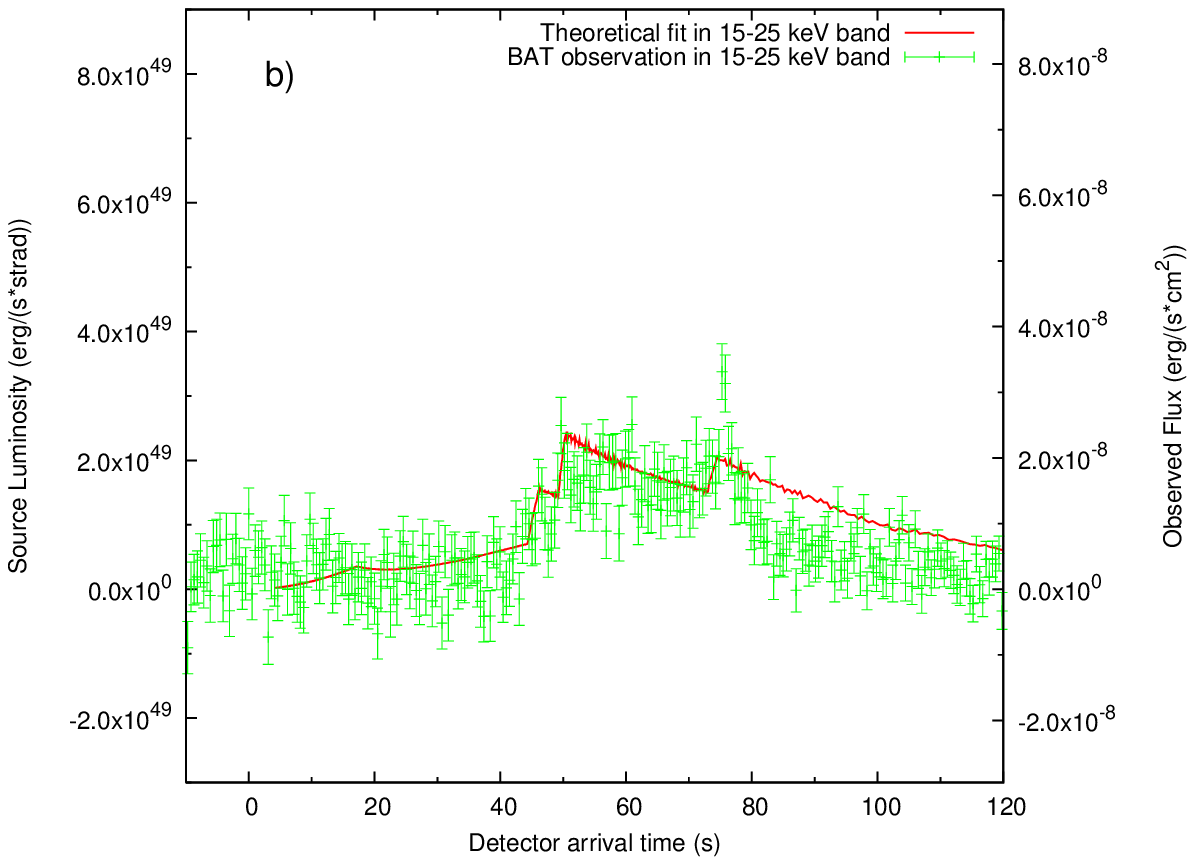}\\
\includegraphics[width=0.5\hsize,clip]{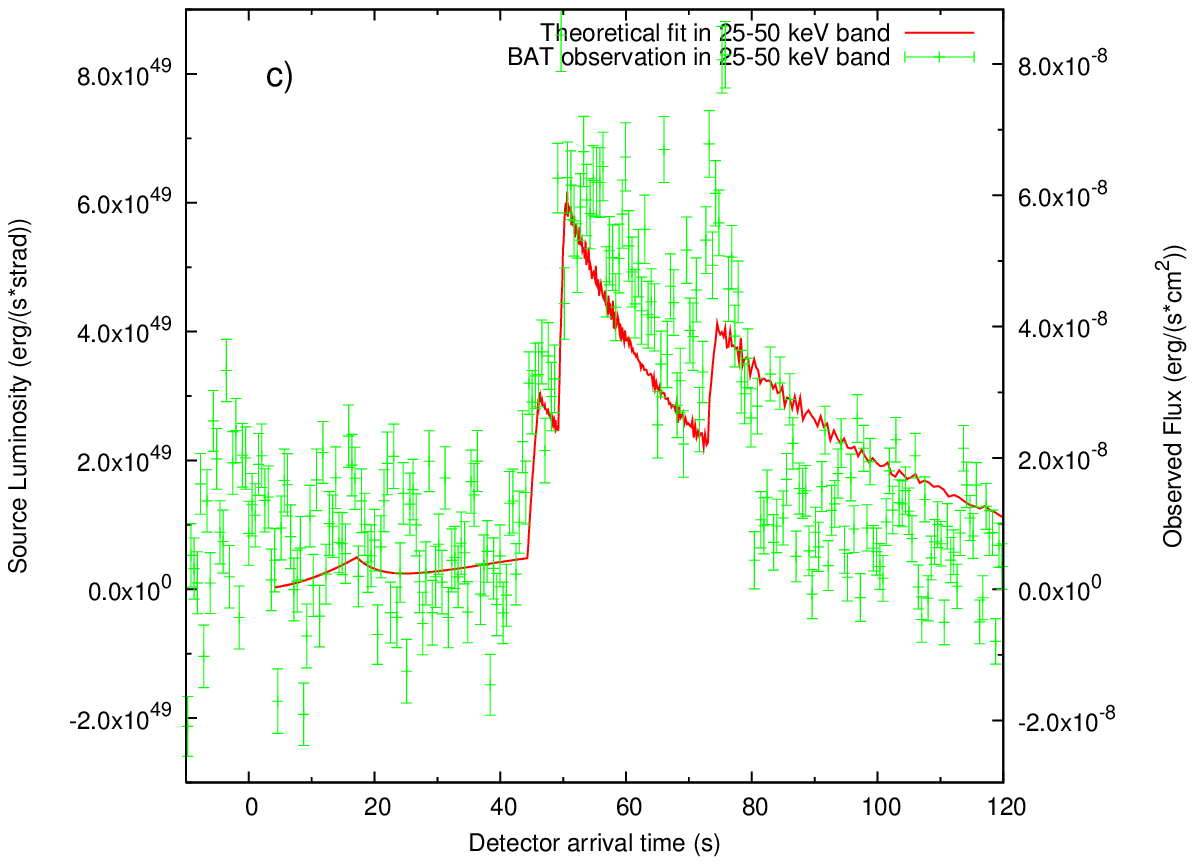}
\includegraphics[width=0.5\hsize,clip]{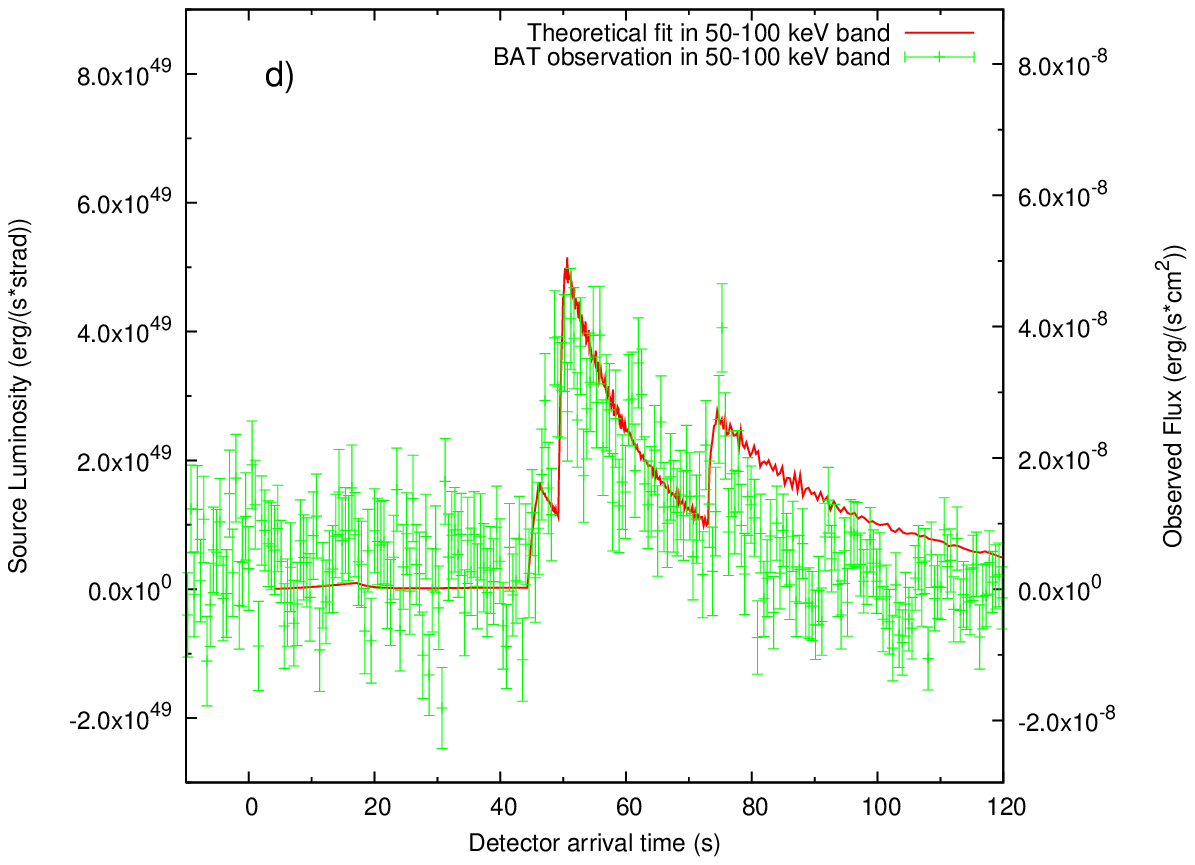}
\caption{Our theoretical fit (red line) of the BAT observations (green points) of GRB 050315 in the $15$--$350$ keV (a), $15$--$25$ keV (b), $25$--$50$ keV (c), $50$--$100$ keV (d) energy bands \citep{va05}. The blue line in panel (a) represents our theoretical prediction for the intensity and temporal position of the P-GRB.}
\label{tot}
\end{figure*}

The verification of the validity of our model has been tested in a variety of other sources, beside GRB 991216 \citep{rubr}, like GRB 980425 \citep{cospar02}, GRB 030329 \citep{mg10grazia}, GRB 031203 \citep{031203}. In all such sources, again, the observational data were available only during the prompt emission and the latest afterglow phases, leaving our theoretical predictions of the in-between evolution untested. Now, thanks to the data provided by the \emph{Swift} satellite, we are finally able to confirm, by direct confrontation with the observational data, our theoretical predictions on the GRB structure with a detailed fit of the complete afterglow light curve of GRB 050315, from the peak, including the ``prompt emission'', all the way to the latest phases without any gap in the observational data.

\section{The fit of the observations}\label{fit}

The best fit of the observational data leads to a total energy of the black hole dyadosphere, generating the $e^\pm$ plasma, $E_{dya} = 1.46\times 10^{53}$ erg \citep[the observational \emph{Swift} $E_{iso}$ is $> 2.62\times 10^{52}$ erg, see][]{va05}, so that the plasma is created between the radii $r_1 = 5.88\times 10^6$ cm and $r_2 = 1.74 \times 10^8$ cm with an initial temperature $T = 2.05 MeV$ and a total number of pairs $N_{e^+e^-} = 7.93\times 10^{57}$. The second parameter of the theory, the amount $M_B$ of baryonic matter in the plasma, is found to be such that $B \equiv M_Bc^2/E_{dya} = 4.55 \times 10^{-3}$. The transparency point and the P-GRB emission occurs then with an initial Lorentz gamma factor of the accelerated baryons $\gamma_\circ = 217.81$ at a distance $r = 1.32 \times 10^{14}$ cm from the black hole.

\subsection{The BAT data}

In Fig. \ref{tot} we represent our theoretical fit of the BAT observations in the three energy channels $15$--$25$ keV, $25$--$50$ keV and $50$--$100$ keV and in the whole $15$--$350$ keV energy band.

In our model the GRB emission starts at the transparency point when the P-GRB is emitted; this instant of time is often different from the moment in which the satellite instrument triggers, due to the fact that sometimes the P-GRB is under the instrumental noise threshold or comparable with it. In order to compare our theoretical predictions with the observations, it is important to estimate and take into account this time shift. In the present case of GRB 050315 it has been observed \citep[see][]{va05} a possible precursor before the trigger. Such a precursor is indeed in agreement with our theoretically predicted P-GRB, both in its isotropic energy emitted (which we theoretically predict to be $E_{P-GRB} = 1.98 \times 10^{51}$ erg) and its temporal separation from the peak of the afterglow (which we theoretically predicted to be $\Delta t^d_a = 51$ s). In Fig. \ref{tot}a the blue line shows our theoretical prediction for the P-GRB in agreement with the observations.

After the P-GRB emission, all the observed radiation is produced by the interaction of the expanding baryonic shell with the interstellar medium. In order to reproduce the complex time variability of the light curve of the prompt emission as well as of the afterglow, we describe the ISM filamentary structure, for simplicity, as a sequence of overdense spherical regions separated by much less dense regions. Such overdense regions are nonhomogeneously filled, leading to an effective emitting area $A_{eff}$ determined by the dimensionless parameter ${\cal R} \equiv A_{eff} / A_{vis}$, where $A_{vis}$ is the expanding baryonic shell visible area \citep[see][for details]{spectr1,fil}. Clearly, in order to describe any detailed structure of the time variability an authentic three dimensional representation of the ISM structure would be needed. However, this finer description would not change the substantial agreement of the model with the observational data. Anyway, in the ``prompt emission'' phase, the small angular size of the source visible area due to the relativistic beaming makes such a spherical approximation an excellent one \citep[see also][for details]{r02}.

The structure of the ``prompt emission'' has been reproduced assuming three overdense spherical ISM regions with width $\Delta$ and density contrast $\Delta n/\langle n\rangle$: we chose for the first region, at $r = 4.15\times 10^{16}$ cm, $\Delta = 1.5\times 10^{15}$ cm and $\Delta n/\langle n\rangle = 5.17$, for the second region, at $r = 4.53\times 10^{16}$ cm, $\Delta = 7.0\times 10^{14}$ cm and $\Delta n/\langle n\rangle = 36.0$ and for the third region, at $r = 5.62\times 10^{16}$ cm, $\Delta = 5.0\times 10^{14}$ cm and $\Delta n/\langle n\rangle = 85.4$. The ISM mean density during this phase is $\left\langle n_{ISM} \right\rangle=0.81$ particles/cm$^3$ and $\left\langle {\cal R} \right\rangle = 1.4 \times 10^{-7}$. With this choice of the density mask we obtain agreement with the observed light curve, as shown in Fig. \ref{tot}. A small discrepancy occurs in coincidence with the last peak: this is due to the fact that at this stage the source visible area due to the relativistic beaming is comparable with the size of the clouds, therefore the spherical shell approximation should be duly modified by a detailed analysis of a full three-dimensional treatment of the ISM filamentary structure. Such a topic is currently under investigation \citep[see also][for details]{r02}. Fig. \ref{tot} shows also the theoretical fit of the light curves in the three BAT energy channels in which the GRB has been detected ($15$--$25$ keV in Fig. \ref{tot}b, $25$--$50$ keV in Fig. \ref{tot}c, $50$--$100$ keV in Fig. \ref{tot}d).

\subsection{The XRT data}

\begin{figure}
\includegraphics[width=\hsize,clip]{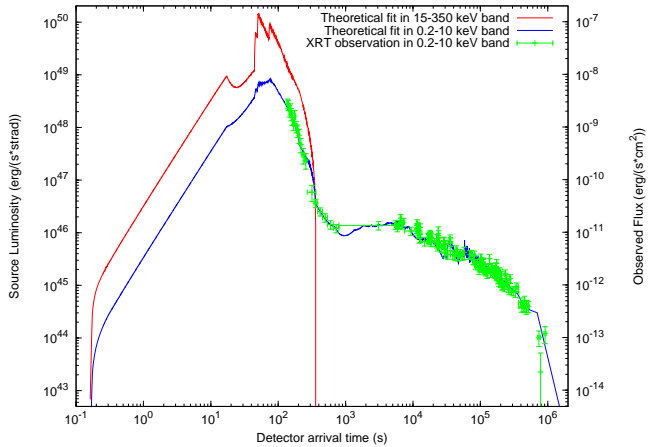}
\caption{Our theoretical fit (blue line) of the XRT observations (green points) of GRB 050315 in the $0.2$--$10$ keV energy band \citep{va05}. The theoretical fit of the BAT observations (see Fig. \ref{tot}a) in the $15$--$350$ keV energy band is also represented (red line).}
\label{global}
\end{figure}

The same analysis can be applied to explain the features of the XRT light curve in the afterglow phase. It has been recently pointed out \citep{nousek} that almost all the GRBs observed by {\em Swift} show a ``canonical behavior'': an initial very steep decay followed by a shallow decay and finally a steeper decay. In order to explain these features many different approaches have been proposed \citep{meszaros,nousek,panaitescu,zhang}. In our treatment these behaviors are automatically described by the same mechanism responsible for the prompt emission described above: the baryonic shell expands in an ISM region, between $r = 9.00\times 10^{16}$ cm and $r = 5.50\times 10^{18}$ cm, which is significantly at lower density ($\left\langle n_{ISM} \right\rangle=4.76 \times 10^{-4}$ particles/cm$^3$, $\left\langle {\cal R} \right\rangle = 7.0 \times 10^{-6}$) then the one corresponding to the prompt emission, and this produces a slower decrease of the velocity of the baryons with a consequent longer duration of the afterglow emission. The initial steep decay of the observed flux is due to the smaller number of collisions with the ISM. In Fig. \ref{global} is represented our theoretical fit of the XRT data, together with the theoretically computed $15$--$350$ keV light curve of Fig. \ref{tot}a (without the BAT observational data to not overwhelm the picture too much).

What is impressive is that no different scenarios need to be advocated in order to explain the features of the light curves: both the prompt and the afterglow emission are just due to the thermal radiation in the comoving frame produced by inelastic collisions with the ISM duly boosted by the relativistic transformations over the EQTSs.

\section{The instantaneous spectrum}\label{spectra}

\begin{figure}
\includegraphics[width=\hsize,clip]{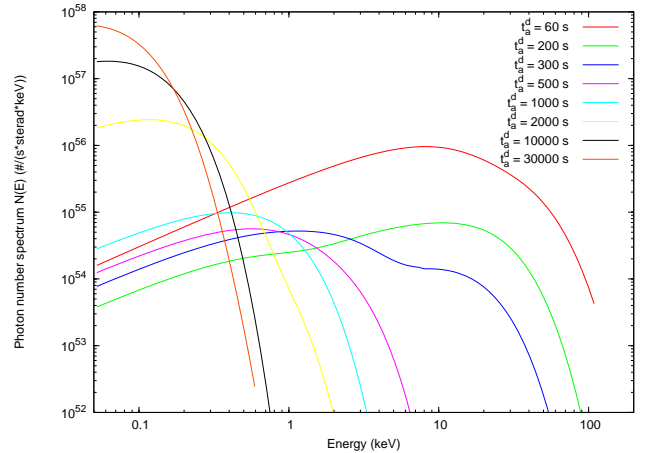}
\caption{Eight theoretically predicted instantaneous photon number spectra $N(E)$ are here represented for different values of the arrival time (colored curves). The hard to soft behavior is confirmed.}
\label{spettro}
\end{figure}

In addition to the the luminosity in fixed energy bands we can derive also the instantaneous photon number spectrum $N(E)$ starting from the same assumptions. In Fig. \ref{spettro} are shown samples of time-resolved spectra for eight different values of the arrival time which cover the whole duration of the event. It is manifest from this picture that, although the spectrum in the co-moving frame of the expanding pulse is thermal, the shape of the final spectrum in the laboratory frame is clearly non thermal. In fact, as explained in \citet{spectr1}, each single instantaneous spectrum is the result of an integration of thousands of thermal spectra over the corresponding EQTS. This calculation produces a non thermal instantaneous spectrum in the observer frame (see Fig. \ref{spettro}).

A distinguishing feature of the GRBs spectra which is also present in these instantaneous spectra is the hard to soft transition during the evolution of the event \citep{cri97,fa00,gcg02}. In fact the peak of the energy distribution $E_p$ drifts monotonically to softer frequencies with time. This feature is linked to the change in the power-law low energy spectral index $\alpha$ \citep{b93}, so the correlation between $\alpha$ and $E_p$ \citep{cri97} is explicitly shown.

It is important to stress that there is no difference in the nature of the spectrum during the prompt and the afterglow phases: the observed energy distribution changes from hard to soft, with continuity, from the ``prompt emission'' all the way to the latest phases of the afterglow.

\section{Conclusions}\label{concl}

Before the \emph{Swift} data, our model could not be directly fully tested. With GRB 050315, for the first time, we have obtained a good match between the observational data and our predicted intensities, in $5$ energy bands, with continuous light curves near the beginning of the GRB event, including the ``prompt emission'', all the way to the latest phases of the afterglow. This certainly supports our model and opens a new phase of using it to identify the astrophysical scenario underlying the GRB phenomena. In particular:
\begin{enumerate}
\item We have demonstrated that the ``prompt emission'' is not necessarily due to the prolonged activity of an ``inner engine'', but corresponds to the emission at the peak of the afterglow.
\item We have a clear theoretical prediction on the total energy emitted in the P-GRB $E_{P-GRB} = 1.98 \times 10^{51}$ erg and on its temporal separation from the peak of the afterglow $\Delta t^d_a = 51$ s. To understand the physics of the inner engine more observational and theoretical attention should be given to the analysis of the P-GRB.
\item We have uniquely identified the basic parameters characterizing the GRB energetics: the total energy of the black hole dyadosphere $E_{dya} = 1.46\times 10^{53}$ erg and the baryon loading parameter $B = 4.55 \times 10^{-3}$.
\item The ``canonical behavior'' in almost all the GRB observed by \emph{Swift}, showing an initial very steep decay followed by a shallow decay and finally a steeper decay, as well as the time structure of the ``prompt emission'' have been related to the fluctuations of the ISM density and of the ${\cal R}$ parameter.
\item The theoretically predicted instantaneous photon number spectrum shows a very clear hard-to-soft behavior continuously and smoothly changing from the ``prompt emission'' all the way to the latest afterglow phases.
\end{enumerate}

Only the first afterglow regime we theoretically predicted, which corresponds to a bolometric luminosity monotonically increasing with the photon detector arrival time, preceding the ``prompt emission'', still remains to be checked by direct observations. We hope in the near future to find an intense enough source, observed by the \emph{Swift} satellite, to verify this still untested theoretical prediction.

As a byproduct of the results presented in this Letter, we can explain one of the long lasting unanswered puzzles of GRBs: the light curves in the ``prompt emission'' show very strong temporal substructures, while they are remarkably smooth in the latest afterglow phases. The explanation follows from three factors: 1) the value of the Lorentz $\gamma$ factor, 2) the EQTS structure and 3) the coincidence of the ``prompt emission'' with the peak of the afterglow. For $\gamma \sim 200$, at the peak of the afterglow, the diameter of the EQTS visible area due to relativistic beaming is small compared to the typical size of an ISM cloud. Consequently, any small inhomogeneity in such a cloud produces a marked variation in the GRB light curve. On the other hand, for $\gamma \to 1$, in the latest afterglow phases, the diameter of the EQTS visible area is much bigger than the typical size of an ISM cloud. Therefore, the observed light curve is a superposition of the contribution of many different clouds and inhomogeneities, which produces on average a much smoother light curve \citep[details in][]{r02,rubr}.

\acknowledgments
We thank P. Banat, G. Chincarini, A. Moretti and S. Vaughan for their help in the analysis of the observational data as well as an anonymous referee for his/her useful considerations.

\end{document}